\documentclass{llncs}
\usepackage{graphicx}
\usepackage{float}
\usepackage{cite}

\usepackage{url}
\usepackage[utf8]{inputenc} 
\usepackage{amsmath}
\usepackage{amsfonts}
\usepackage[usenames,dvipsnames]{color}
\begin{document}

\title{TimeMachine: Entity-centric Search and Visualization of News Archives}
\author{Pedro Saleiro$^{1,2}$, Jorge Teixeira$^{1,2,3}$, Carlos Soares$^{1,2, 4}$, Eug{\'e}nio Oliveira$^{1,2,3}$}

\institute{Labs Sapo UP$^{1}$, DEI-FEUP$^{2}$, LIACC$^{3}$, INESC-TEC$^{4}$,\\University of Porto,\\Rua Dr. Roberto Frias, s/n, Porto, Portugal}
\maketitle

\begin{abstract}

We present a dynamic web tool that allows interactive search and visualization of large news archives using an entity-centric approach.  Users are able to search entities using keyword phrases expressing news stories or events and the system retrieves the most relevant entities to the user query based on automatically extracted and indexed entity profiles. From the computational journalism perspective, TimeMachine allows users to explore media content through time using automatic identification of entity names, jobs, quotations and relations between entities from co-occurrences networks extracted from the news articles. TimeMachine demo is available at \textcolor{blue}{http://maquinadotempo.sapo.pt/}.

\end{abstract}

\section{Introduction}

Online publication of news articles has become a standard behavior of news outlets, while the public joined the movement either using desktop or mobile terminals. The resulting setup consists of a cooperative dialog between news outlets and the public at large.  Latest events are covered and commented by both parties in a continuous basis through the social networks, such as Twitter. At the same time, it is necessary to convey how story elements are developed over time and to integrate the story in the larger context. This is extremely challenging when journalists have to deal with news archives that are growing everyday in a thousands scale. Never before has computation been so tightly connected with the practice of journalism. In recent years, computer science community have researched \cite{demartini2010taer, matthews2010searching, balog2009sahara, alonso2010time, saleiro2015popmine, Teixeira2011, Sarmento2009, Abreu2015}  and developed\footnote{NewsExplorer (IBM Watson): http://ibm.co/1OsBO1a} new ways of processing and exploring news archives to help journalists perceiving news content with an enhanced perspective.

TimeMachine, as a computational journalism tool, brings together a set of Natural Language Processing, Text Mining and Information Retrieval technologies to automatically extract and index entity related knowledge from the news articles \cite{saleiro2013popstar, saleiro2013piaf, saleiro2015popmine, teixeira2011bootstrapping, Teixeira2011, Sarmento2009,Abreu2015 }. It allows users to issue queries containing keywords and phrases about news stories or events, and retrieves the most relevant entities mentioned in the news articles through time. TimeMachine provides readable and user friendly insights and visual perspective of news stories and entities evolution along time, by presenting co-occurrences networks of public personalities mentioned on news, following a force atlas algorithm \cite{jacomy2014forceatlas2} for the interactive and real-time clustering of entities.

\section{News Processing Pipeline}

The news processing pipeline, depicted in Figure 1, starts with a news cleaning module which performes the boilerplate removal from the news raw files (HTML/XML). Once the news content is processed we apply the NERD module which recognizes entity mentions and disambiguates each mention to an entity using a set of heuristics tailored for news, such as job descriptors (e.g. ``Barack Obama, president of USA'') and linguistic patterns well defined for the journalistic text style. 
We use a bootstrap approach to train the NER system \cite{teixeira2011bootstrapping}. Our method starts by annotating persons names on a dataset of 50,000 news items. This is performed using a simple dictionary-based approach. Using such training set we build a classification model based on Conditional Random Fields (CRF). We then use the inferred classification model to perform additional annotations of the initial seed corpus, which is then used for training a new classification model. This cycle is repeated until the NER model stabilizes.
\begin{figure}[h!]
    \centering
    \includegraphics[width=0.8\columnwidth]{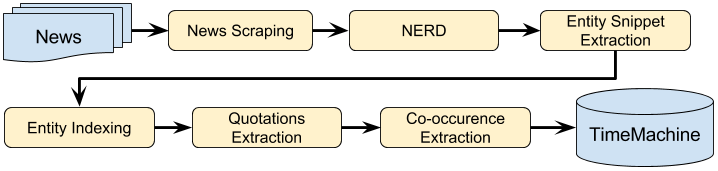}
    \caption{News processing pipeline.}
    \label{fig:architecture}
\end{figure}
The entity snippet extraction consists of collecting sentences containing mentions to a given entity. All snippets are concatenated generating an entity document, which will be indexed in the entity index. The entity index represents the frequency of co-occurrence of each entity with each term that it occurs with in the news. Therefore, by relying on the redundancy of news terms and phrases associated with an entity we are able to retrieve the most relevant entity to a given input keyword or phrase query. As we also index the snippet datetime it is possible to filter query results based on a time span. For instance, the keyword ``corruption'' might retrieve a different entity list results in different time periods. Quotations are typically short and very informative sentences, which may directly or indirectly quote a given entity. Quotations are automatically extracted (refer to "Quotations Extraction" module) using linguistic patterns, thus enriching the information extracted for each entity.
Finally, once we have all mentioned entities in a given news articles we extract entity tuples representing co-occurrences of entities in a given news article and update the entity graph by incrementing the number of occurrences of a node (entity) and creating/incrementing the number of occurrences of the edge (relation) between any two mentions.

\section{Demonstration}
The setup for demonstration uses a news archive of Portuguese news. It comprises two different datasets: a repository from the main Portuguese news agency (1990-2010), and a stream of online articles provided by the main web portal in Portugal (SAPO) which aggregates news articles from 50 online newspapers. By the time of writing this paper, the total number of news articles used in this demonstration comprises over 12 million news articles. The system is working on a daily basis, processing articles as they are collected from the news stream.
TimeMachine allows users to explore its news archive through an entity search box or by selecting a specific date. Both options are available on the website homepage and in the top bar on every page. There are a set of ``stories'' recommendations on the homepage suited for first time visitors. The entity search box is designed to be the main entry point to the website as it is connected to the entity retrieval module of TimeMachine. 
\begin{figure}[H]
    \centering
    \includegraphics[width=0.9\textwidth]{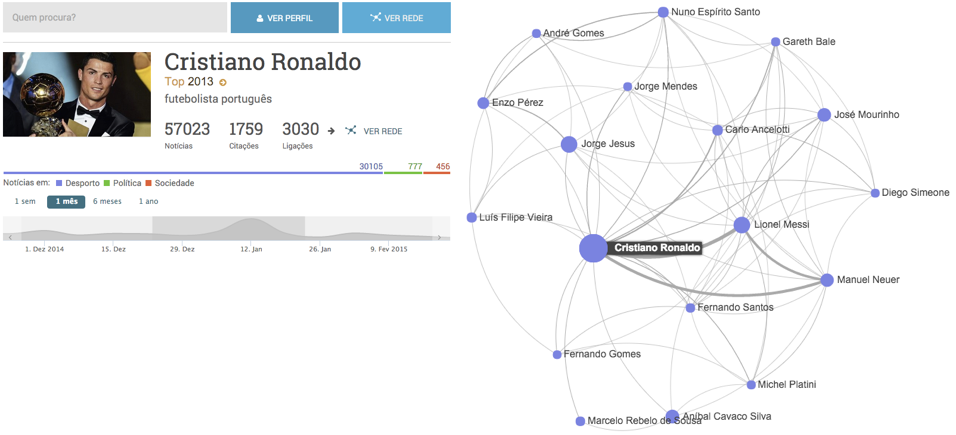}
    \caption{Cristiano Ronaldo page headline (left) and egocentric network (right).}
    \label{fig:architecture}
\end{figure}
Users may search for surface names of entities (e.g. ``Cristiano Ronaldo'') if they know which entities they are interested to explore in the news, although the most powerful queries are the ones containing keywords or phrases describing topics or news stories, such as ``eurozone crisis'' or  ``ballon d'or nominees''. When selecting an entity from the ranked list of results, users access the entity profile page which containing a set of automatically extracted entity specific data: name, profession, a set of news articles, quotations from the entity and related entities. Figure 2, left side, represents an example of the entity profile headline. The entity timeline allows users to navigate entity specific data through time. By selecting a specific period, different news articles, quotations and related entities are retrieved.  Furthermore, users have the option of ``view network'' which consists in a interactive network depicting connections among entities mentioned in news articles for the selected time span. This visualization is depicted in Figure 2, right side, and it is implemented using the graph drawing library Sigma JS, together with "Force Atlas" algorithm for the clustering of entities. Nodes consist of entities and edges represent a co-occurrence of mentioned entities in the same news article. The size of the nodes and the width of edges is proportional to the number of mentions and co-occurrences, respectively. Different node colors represent specific news topics where entities were mentioned. By selecting a date interval on the homepage, instead of issuing a query, users get a global interactive network of mentions and co-occurrences of the most frequent entities mentioned in the news articles for the selected period of time. 

As future work we plan to enhance TimeMachine with semantic extraction and retrieval of relations between mentioned entities.

\bibliographystyle{unsrt}
\bibliography{refs}  

\end{document}